\documentclass[twocolumn,aps,pre,amsmath,nofootinbib,longbibliography]{revtex4-1}
\usepackage{float}
\usepackage{color}
\usepackage{graphicx}
\usepackage{ulem}
\usepackage{epstopdf}
\usepackage{amsfonts}
\usepackage{amsmath}
\usepackage{mathtools}
\usepackage{bbold}
\usepackage[toc,page]{appendix}
\usepackage[mathscr]{eucal}

\newcommand{\beginsupplement}{%
        \setcounter{table}{0}
        \renewcommand{\thetable}{S\arabic{table}}%
        \setcounter{figure}{0}
        \renewcommand{\thefigure}{S\arabic{figure}}%
        \setcounter{equation}{0}
        \renewcommand{\theequation}{S\arabic{equation}}%
     }

\newcommand{\eC}{\overline{\mathbf{C}}}
\newcommand{\eCr}{\overline{\mathbf{C}}_{[{\rm r},{\rm r}]}}

\newcommand{\x}{\mathbf{x}}
\newcommand{\xr}{\mathbf{x}_{\rm r}}

\newcommand{\A}{\mathbf{A}}

\newcommand{\C}{\mathbf{C}}
\newcommand{\Cr}{\mathbf{C}_{[{\rm r},{\rm r}]}}
\newcommand{\D}{\mathbf{D}}
\newcommand{\Dr}{\mathbf{D}_{[{\rm r},{\rm r}]}}

\newcommand{\Fr}{\mathbf{f}_{\rm r}}

\newcommand{\J}{\mathbf{j}}

\newcommand{\M}{\rm M}
\newcommand{\Dip}{\rm D}

\newcommand{\pdx}{\widetilde{\partial}_{x}^2}
\newcommand{\pdy}{\widetilde{\partial}_{y}^2}
\newcommand{\pdc}{\partial ^2}

\newcommand{\z}{\mathbf{z}}
\newcommand{\iv}{\mathbf{i}}
\newcommand{\jv}{\mathbf{j}}

\newcommand\chout{\bgroup\markoverwith{\textcolor{red}{\rule[0.5ex]{2pt}{1.0pt}}}\ULon}
\newcommand\fedout{\bgroup\markoverwith{\textcolor{blue}{\rule[0.5ex]{2pt}{1.0pt}}}\ULon}
\newcommand\grzout{\bgroup\markoverwith{\textcolor{green}{\rule[0.5ex]{2pt}{1.0pt}}}\ULon}

\DeclareMathOperator{\Tr}{Tr}
\DeclareSymbolFont{newfont}{OML}{cmm}{m}{it}
\DeclareMathSymbol{\Varrho}{3}{newfont}{37}
 
\begin{document}

\title{Mesoscopic non-equilibrium measures can reveal intrinsic features of the active driving}
\author{Federica Mura}
\affiliation{Arnold-Sommerfeld-Center for Theoretical Physics and Center for
  NanoScience, Ludwig-Maximilians-Universit\"at M\"unchen,
   D-80333 M\"unchen, Germany.}
\author{Grzegorz Gradziuk}
\affiliation{Arnold-Sommerfeld-Center for Theoretical Physics and Center for
  NanoScience, Ludwig-Maximilians-Universit\"at M\"unchen,
   D-80333 M\"unchen, Germany.}
\author{Chase P. Broedersz}
\email{C.broedersz@lmu.de}
\affiliation{Arnold-Sommerfeld-Center for Theoretical Physics and Center for
  NanoScience, Ludwig-Maximilians-Universit\"at M\"unchen,
   D-80333 M\"unchen, Germany.}

\pacs{}
\date{\today}

\begin{abstract}
Biological assemblies such as chromosomes, membranes, and the cytoskeleton are driven out of equilibrium at the nanoscale by enzymatic activity and molecular motors. Similar non-equilibrium dynamics can be realized in synthetic systems, such as chemically fueled colloidal particles. Characterizing the stochastic non-equilibrium dynamics of such active soft assemblies still remains a challenge. Recently, new non-invasive approaches have been proposed to determine non-equilibrium behavior, which are based on detecting broken detailed balance in the stochastic trajectories of several coordinates of the system. Inspired by the method of two-point microrheology, in which the equilibrium fluctuations of a pair of probe particles reveal the viscoelastic response of an equilibrium system, here we investigate whether we can extend such an approach to non-equilibrium assemblies: can one extract information on the nature of the active driving in a system from the analysis of a two-point non-equilibrium measure?  We address this question theoretically in the context of a class of elastic systems, driven out of equilibrium by a spatially heterogeneous stochastic internal driving. We consider several scenarios for the spatial features of the internal driving that may be relevant in biological and synthetic systems, and investigate how such features of the active noise may be reflected in the long-range scaling behavior of two-point non-equilibrium measures. 

\end{abstract}
\maketitle
\noindent 

\section{Introduction}

Active matter theories aim to provide a physical description for systems intrinsically out of thermal equilibrium. A prominent collection of such systems is classified as soft biological materials, with typical examples as  tissue, membranes and cytoskeletal structures~\cite{Fodorreview, Needleman2017b, Gnesottoreview,mackintosh2010}. These soft materials can be easily deformed by thermally driven stresses. However, temperature is not the only source of fluctuations in these systems: additional athermal fluctuations are generated at the molecular scale by enzymes that drive the system out of thermal equilibrium. Examples of soft non-equilibrium materials are also found in artificial and biomimetic systems, such as chemically fueled synthetic fibers and crystals of active colloidal particles~\cite{Grotsch2018,Palacci2013,Bertrand2012}.
In all these systems, traces of non-equilibrium may propagate from molecular to larger scales, with striking examples in biology, such as the mitotic spindle and protein pattern formation~\cite{Frey2017,Brugues2014,Gadde2004}. However, non-equilibrium may also manifests as random fluctuations, seemingly indistinguishable from simple thermal motion. For instance, active dynamics were observed in the fluctuations of biological assemblies, such as chromosomes~\cite{Weber}, tissue\cite{Fodor2015}, membranes~\cite{Turlier2016,Betz2009,Ben-Isaac2011} and cytoplasm~\cite{Mizuno2007,Brangwynne2008,Guo2014, Fodor2015a}. This active dynamics can affect the  macroscopic mechanical properties of soft materials and a systematic non-equilibrium characterization could help guide the development of engineered biomaterials~\cite{ Needleman2017b, Broedersz2011a, Agarwal2009, Koenderink2009}.  While we have a comprehensive toolset to measure the equilibrium response of thermal soft materials, it still remains an outstanding challenge to characterize the stochastic non-equilibrium dynamics of active soft assemblies.

A well-established approach to quantify non-equilibrium is based on the violation of the fluctuation-dissipation theorem (FDT). The idea of this approach is to compare the fluctuation spectrum of a probe particle with the associated response function to investigate if these two quantities obey the FDT. This method has been used both in \textit{in vivo} biological assemblies and \textit{in vitro} reconstituted networks~\cite{Mizuno2007,Guo2014,Fodor2015a,Betz2009,Julicherreview}. However, such an approach requires a measurement of the system's response function, which may be technically difficult especially in living systems. 

Recently, new non-invasive approaches have been proposed, based on the detection of the irreversibility of stochastic trajectories. Such irreversibility can be expressed, for instance, in the form of broken detailed balance~\cite{Battle2016, Gladrow2016, Gladrow2017,Mura2018, Gradziuk2019} or in terms of the entropy production rate~\cite{Mura2018, Frishman2018, Li2019a,Seara2018,Seifertreview}.
Broken detailed balance can be determined from circulating currents in the coordinates space of pairs of mesoscopic degrees of freedom. Frequently used measures to quantify circulating currents in phase space are the area enclosing rate and the cycling frequency of the stochastic trajectories~\cite{Ghanta2017,Gonzalez2018,Gladrow2016}. 
These measures are closely related to the entropy production rate~\cite{Mura2018, Gradziuk2019}. 

The interpretation of the results of various non-equilibrium measurements can be aided by considering concrete models for active systems. 
Recently, we considered a simple model of driven elastic assemblies consisting of a bead-spring network~\cite{Mura2018,Gradziuk2019} where the beads can experience both thermal and active fluctuations. Using this model, we estimated the area enclosing rate and the cycling frequency from the trajectories of two probe particles. On average such non-equilibrium measures exhibit a power law behavior as a function of the distance between the probe particles. Inspired by the approach of two-point microrheology, in which the fluctuations of a pair of probe particles reveal the viscoelastic response of an equilibrium system~\cite{Levine2000}, here we investigate whether we can extend such an approach to non-equilibrium assemblies: can one extract information on the nature of the active driving in an elastic assembly, simply from the analysis of a two-point non-equilibrium measure?

We consider this question in the context of a class of elastic systems with stochastic internal driving. This internal driving can be described as a stochastic process with specific statistical properties that characterize their spatial and temporal features. Here we focus on the spatial properties of the stochastic internal driving and investigate several scenarios that may be relevant in biological systems. For instance, the internal driving can be implemented as a heterogeneous distribution of spatially and temporally uncorrelated stochastic forces. This description may be adequate to represent the enhanced diffusion experienced within certain regions of a cellular environment due to catalytic enzymes~\cite{Jee2018,Golestanian2015,Riedel2015}. However, the sources of activity in a biological environment can take a variety of forms, including contributions from chemophoresis~\cite{Sugawara2011} or molecular motors. For instance, force-generation by molecular motors such as myosin can be introduced as force dipoles randomly distributed over the network~\cite{Chen2011,Moller2016,Broedersz2011a,Joanny2009}. Furthermore, the intracellular organization of enzymes and molecular motors may also result in long-range correlation of the activities~\cite{Sutherland2009,Hachet2011,Schmitt2017}.  

In this work we investigate how such intrinsic spatial features of the active noise may be reflected in the long-range scaling behavior of two-point non-equilibrium measures.
We employ a model of internally driven elastic networks to describe biological assemblies. To consider a general scenario, we describe internal activity with stochastic forces acting either as monopoles or as dipoles. Within this framework we show analytically and numerically how the scaling behavior of the cycling frequencies and area enclosing rates depends on the parameters that characterize the active noise, i.e  intensity and density of the monopole- and dipole-like activities. Finally, we show how our framework can be extended to account for spatial correlations in the intensities of the noise. 
Our results give insights into possible methods of quantifying non-equilibrium in biological assemblies and, more specifically, into how an experimental observation of a particular scaling behavior of non-equilibrium measures can give access to the properties of the active driving.
\section{Model}
\begin{figure}
  \includegraphics[width=8.5cm]{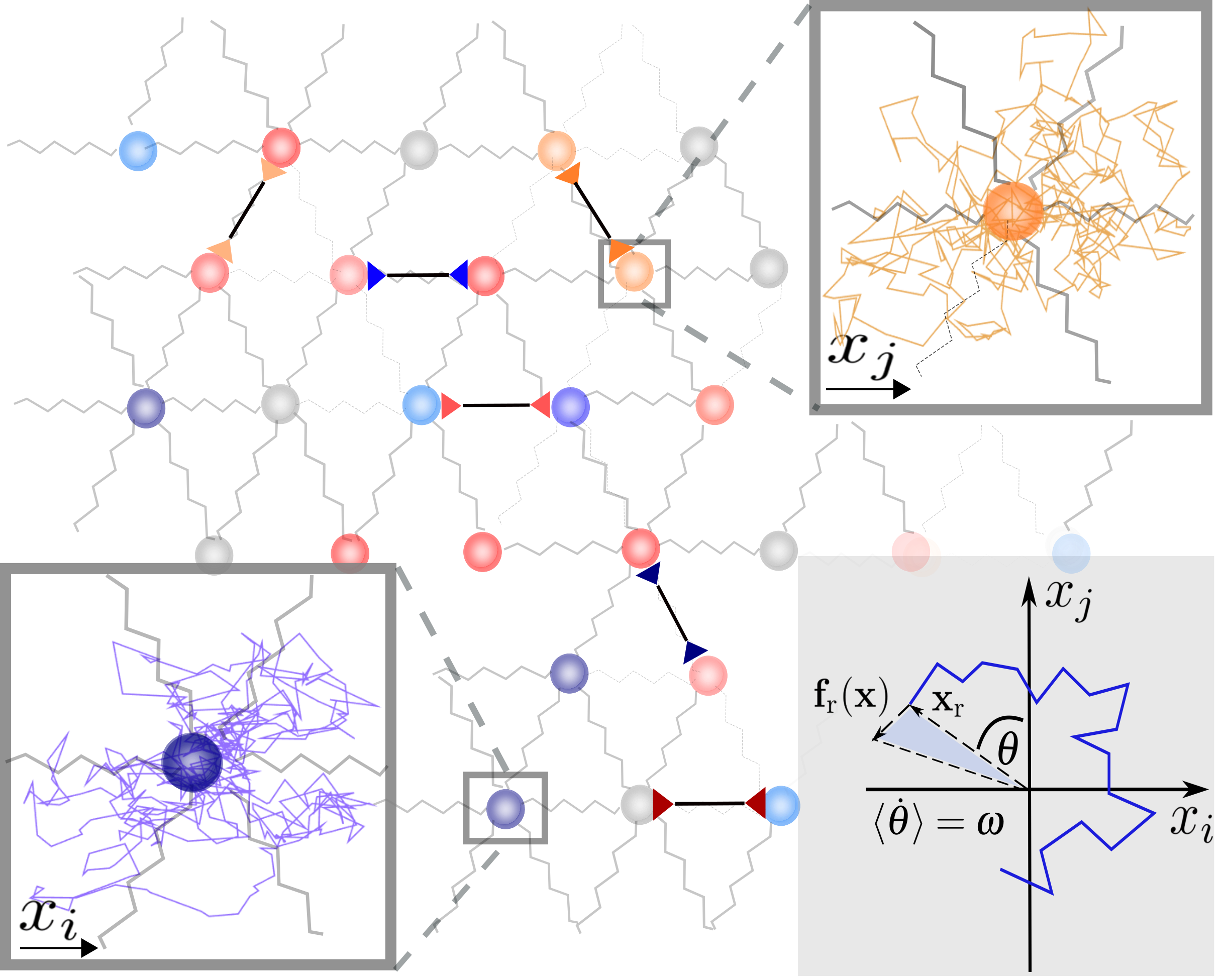}

  \caption{A) Schematic of the elastic network with monopole and dipole forces. The colors of the beads indicate the intensities of the monopole-like activity, the colors of the arrows indicate the intensity of the dipole-like activity. Inset on the lower right: schematic of a trajectory in the $x_i$ and $x_j$ coordinate space. The indicated non-equilibrium measures are: cycling frequency $\omega$, and area enclosing rate obtained upon averaging $(\Fr \times \xr)/2\gamma$ (light blue area) in phase space. 
}
    \label{fig:Fig1_dipolescartoon}
\end{figure}
To establish a relation between mesoscopic non-equilibrium measures and the internal activity, we consider a model for an elastic assembly driven by stochastic activity~\cite{Mura2018, Harder2014,Mallory2015,Kaiser2015}. Our model consists of a network of $N$ beads connected by springs of elastic constant $k$. The network is immersed in a thermal bath, characterized by a friction coefficient $\gamma$ and temperature $T$. Within the cellular environment there are several sources of athermal fluctuations, such as enhanced diffusion of catalytic enzymes~\cite{Jee2018,Golestanian2015,Riedel2015}, or the action of molecular motors. While the former could be modeled as thermal-like monopole force fluctuations with enhanced variance acting on the beads, the latter is better described as stochastic force dipoles~\cite{Chen2011,Moller2016,Broedersz2011a,Joanny2009}. 

To describe a general scenario, we generate the active driving in our network as Gaussian white noise of either a monopole or a dipole nature, with respective densities  $\rho ^{\M}$ and $\rho ^{\Dip}$ (Fig.~\ref{fig:Fig1_dipolescartoon}). By modeling the active forces as a white noise process, we assume a time scale separation between the dynamics of the microscopic active forces and the shortest relaxation times of the system.

We denote the displacements of the beads relative to their rest positions by the vector $\x = \{x_1,\ldots,x_{Nd}\}$, where $d$ indicates the dimensionality of the system.
The overdamped equation of motion for the displacement of the $i_{\rm th}$ degree of freedom in the lattice reads
\begin{equation}
\frac{dx_i (t)}{dt}=a_{ij}x_j(t) + \eta ^T _i(t) + b^{\M} _i\eta _i ^{\M}(t) +\sum _{k \in {\rm nn}} {b}_{ik}^{\Dip} \eta _{ik} ^{\Dip}(t),
\end{equation}
where $a_{ij}$ are the entries of the elastic interaction matrix $\A$, divided by the friction coefficient $\gamma$.
The thermal noise is described by the standard fluctuation-dissipation relation 
\begin{equation}
\langle \eta_i ^{T}(t) \eta_j ^{T}(t') \rangle=\frac{2k_{\rm B}T}{\gamma} \delta _{i,j}\delta(t-t'),
\end{equation} 
where $k_{\rm B}$ indicates the Boltzmann constant.
The coefficients $b^{\M} _i $ and $b^{\Dip} _{ik} $ are introduced to describe the presence of monopole and dipole active noise; they are time-independent random variables such that $b^{{\M}/{\Dip}} _{i/ik} \in \{0,1\}$, ${b}_{ik}^{\Dip}={b}_{ki}^{\Dip}$, with probability distribution $P\left(b^{{\M}/{\Dip}} _{i/ik}= 1\right)=\rho ^{{\M}/{\Dip}}$. The sum $\sum _{k \in {\rm nn}}$ runs only over nearest-neighbor beads.
Finally, the stochastic variables for monopole forces $\eta ^{\M} (t) $ and dipole forces $\eta ^{\Dip} (t)$ are characterized by
\begin{align}
\begin{split}
 \langle \eta_i ^{\M}(t) \eta_j ^{\M}(t') \rangle&=\frac{k_{\rm B}}{\gamma} \delta _{i,j}\alpha _i ^{\M}\delta(t-t')\\
 \langle \eta_{ij} ^{\Dip}(t) \eta_{kl} ^{D}(t') \rangle&=\frac{k_{\rm B}}{\gamma} \left(\delta _{ij,kl}- \delta _{ij,lk} \right)\alpha _{ij} ^{\Dip}\delta(t-t').
\end{split}
\end{align}
Here we indicate with $\alpha _i ^{\M}$ and $\alpha _{ij} ^{\Dip}$ the respective amplitudes  of the monopole force acting on the $i_{\rm th}$ coordinate and of the dipole force acting between the $i_{\rm th}$ and $j_{\rm th}$ coordinates.  We factored out the term $(k_{\rm B}/ \gamma )$ for notational convenience.

This simple model admits a Fokker Plank description for the evolution of the probability distribution $p(\x,t)$ of $\x$ at time $t$
\begin{equation}
\partial_t p(\x,t)=-\nabla \cdot (\A\x-\D\nabla)p(\x,t)\coloneqq-\nabla \cdot \J(\x,t), 
\label{eq:fokker}
\end{equation}
where $\J(\x,t)$ is the probability current density, and $\D$ is the diffusion matrix. The steady-state solution of Eq.~\eqref{eq:fokker} is a Gaussian distribution $p(\x) \sim \exp[-\frac{1}{2}\x^T\C^{-1}\x]$, with covariance matrix $\C$, satisfying the Lyapunov equation: 
\begin{equation}
 \A\C + \C\A^{\rm T}= -2\D.
\label{eq:Lyapunov}
\end{equation}
The diffusion matrix can be  expressed as $\D= \bar{\D}+ \D ^*$, where $\bar{\D}$  is a diagonal matrix with entries  $\bar{d}_{ij}=\delta_{ij}k_{\rm B}T/\gamma$, representing the thermal noise contributions, and $\D ^*$ is the non-equilibrium part of the diffusion matrix which contains information on the activities. The presence of dipole forces gives rise to anticorrelations between neighboring beads and therefore to non-zero off-diagonals terms in $\D ^*$, as will be discussed in Sec.~(\ref{results}).

\subsection{Two-point non-equilibrium measures}
Our main goal is to connect a direct measure of non-equilibrium to the properties of the active driving. With this in mind, the first step is to define a two-point non-equilibrium measure which can be estimated from the trajectories of pairs of probe particles in the system.

Under steady-state conditions Eq.~\eqref{eq:fokker} reduces to $\nabla~\cdot~\J(\x,t)~=~0$.
When $\mathbf{j}\ne 0$ the system is out of equilibrium, and exhibits on average a circulation in phase space. Such circulation may emerge also in the reduced subspace of a pair of degrees of freedom $\xr=\{x_i,x_j\}$. These reduced subspaces are more easily accessible experimentally as compared to the full set of degrees of freedom. For this reason we restrict our non-equilibrium measures to these two-dimensional subspaces.

As a first measure of circulation, we use the average area enclosing rates $\mathscr{A} _{ij}$ of the trajectory in the reduced subspace of the coordinates $x_i$ and $x_j$, as illustrated in Fig.~\ref{fig:Fig1_dipolescartoon}.
For an overdamped system, for which the velocity is proportional to the force, this quantity can be expressed as~\cite{Gradziuk2019}:
\begin{equation}
\mathscr{A} _{ij}= \frac{1}{2\gamma}\langle \xr \times \Fr(\x) \rangle 
\label{eq:AERtorque}
\end{equation}
where $\Fr(\x)=\{f_i,f_j\}$ is the vector of forces acting respectively on the coordinate $i$ and $j$. By replacing $\gamma ^{-1}\langle \xr \times \Fr(\x) \rangle =\gamma ^{-1}\langle x_i f_j(\x) - x_jf_i(\x) \rangle= (\C\A^{\rm T}- \A\C)_{ij}$ we obtain~\cite{Gradziuk2019,Ghanta2017,Gonzalez2018}.
\begin{equation}
\mathscr{A} _{ij}= \frac{1}{2} (\C\A^{\rm T}- \A\C)_{ij}, 
\label{eq:AER}
\end{equation}
This non-equilibrium measure turns out to be closely related to the cycling frequency-- the rate at which the trajectory revolves in the coordinates space: 
\begin{equation}
\omega_{ij}= \frac{1}{2}\frac{(\C\A^{\rm T}- \A\C)_{ij}}{\sqrt{\det(\Cr)}},
\label{eq:frequency}
\end{equation}
where $\Cr$ is a $[2 \times 2]$ matrix with entries $\Cr=\{\{c_{ii},c_{ij}\},\{c_{ji},c_{jj}\}\}$.
Unlike the area enclosing rate, the cycling frequency is invariant under an orientation preserving change of basis. This invariance is ensured by the factor $\sqrt{\det(\Cr)}$ in Eq.~\eqref{eq:frequency}. Furthermore, the cycling frequency is informative of the partial rate of entropy produced in the reduced subspace of the pair of observed degrees of freedom, through the expression $\Pi ^{(2)}=k_B\omega^2_{ij} \Tr (\Cr\Dr^{-1})$ ~\cite{Mura2018}.

Both these non-equilibrium measures display on average a power law behavior as function of the distance $r$ between the tracked particles~\cite{Mura2018,Gradziuk2019}.
However, different properties of the active noise in the system may give rise to different functional forms of the scaling behavior. 
In Sec.~(\ref{results}) we investigate this matter, and in particular we aim to connect the scaling of these experimentally accessible non-equilibrium measures to the intrinsic properties of the internal driving. 

\section{Results}
\label{results}
 
In this section we analytically and numerically study the spatial scaling behaviors for $\langle \mathscr{A} ^2 (r)\rangle _{\alpha}$ and $\langle \omega^2 (r)\rangle _{\alpha}$, the squared area enclosing rate and cycling frequency between two tracer beads at distance $r$, averaged over the distribution of activities  $ \alpha _{i/ij} ^{M/D}$. Here we indicate  with $\langle \ldots \rangle _{\alpha}$ the ensemble average over the activities, which for a large enough system can be obtained as a spatial average over the network~\cite{Mura2018}. Our analytical expressions for the scaling laws of $\langle \mathscr{A} ^2 (r)\rangle _{\alpha}$ and $\langle \omega^2 (r)\rangle _{\alpha}$ provide a direct connection between these non-equilibrium measures and the properties of the active noise, such as the densities and activity intensities of monopoles and dipoles forces. The validity of our analytical results obtained in this section will be tested by making a direct comparison with numerical solutions of Eq.~\eqref{eq:AER} and Eq.~\eqref{eq:frequency}.  In Sec.~(\ref{sec1d}) we focus for simplicity on a one-dimensional chain, and in Sec.~(\ref{sec2d}) we show an extension to a two-dimensional network. Finally, in Sec.~(\ref{sec_corr}) we discuss how the scaling law is affected by the presence of finite spatial correlation of the intensities of the active noise.
\subsection{One-dimensional chain }
\label{sec1d}
\begin{figure}[t]
\centering
  \includegraphics[width=9cm]{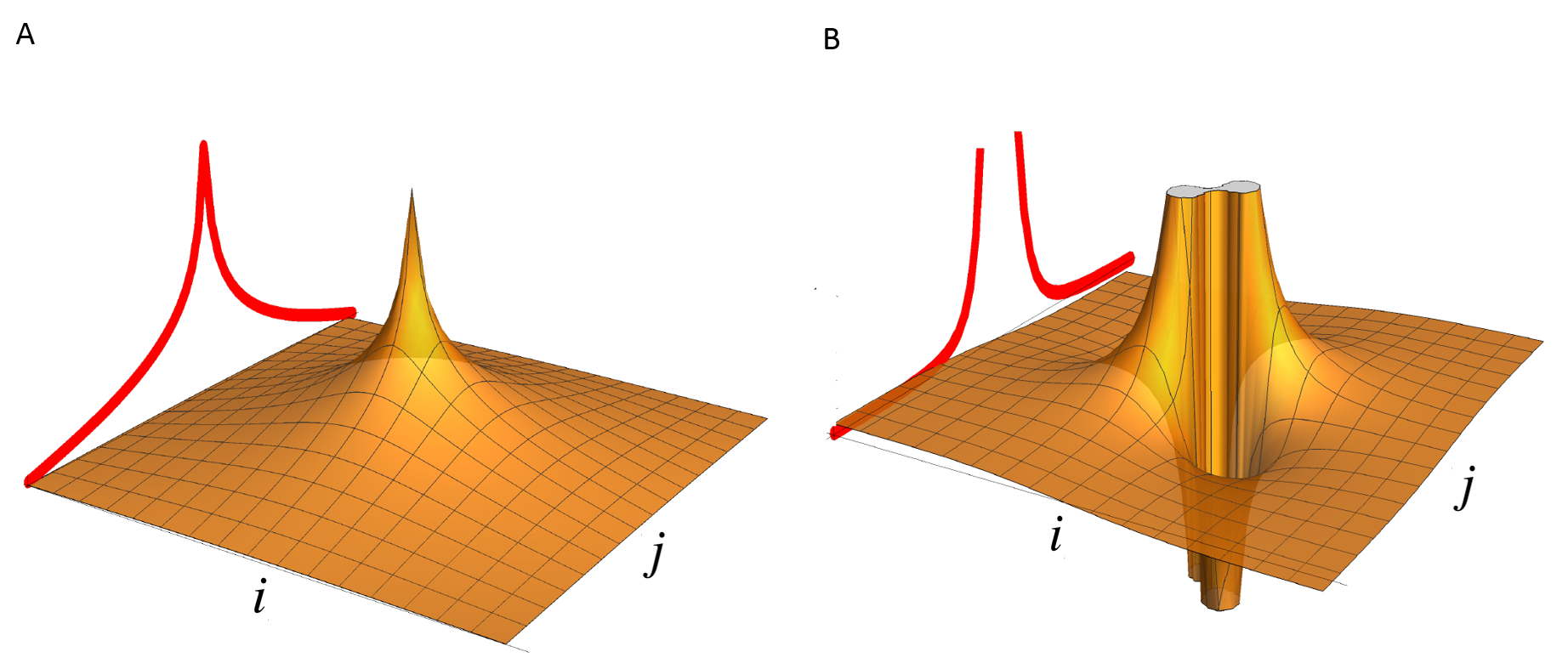}

  \caption{A) Covariance matrix of single monopole source and its projection on the $jz$ plane B) Covariance matrix of single dipole source and its projection on the $jz$ plane.}
    \label{fig:covsing}
\end{figure}
We consider here the simple yet instructive case of a one-dimensional chain. Using a continuous approximation, we find a solution of Eq.~\eqref{eq:Lyapunov} for the covariance matrix and derive the scaling laws for the non-equilibrium measures  $\langle \mathscr{A} ^2 \rangle _{\alpha} $ and $\langle \omega^2 \rangle _{\alpha}$ as function of the distance $r$ between the tracer particles.

As a first step, we determine the form of the non-equilibrium part of the diffusion matrix $\D$ appearing in Eq.~\eqref{eq:Lyapunov}. 
The simple configuration of a one-dimensional case allows us to replace the double index of the dipoles amplitudes $\{ \alpha ^{\Dip}_{12},\alpha ^{\Dip}_{23},\ldots,\alpha ^{\Dip}_{i,i+1},\ldots,\alpha ^{\Dip}_{N-1,N}\}$ with a single index running over the pairs of nearest-neighbor beads $\{ \alpha ^{\Dip}_{1}, \alpha ^{\Dip}_{2},\ldots, \alpha ^{\Dip}_{i}, \ldots, \alpha ^{\Dip}_{N-1}\}$. The activities amplitudes $\{ \alpha _i ^{\M / \Dip} \}$ are sampled independently from a distribution $p^{M/D}_{\alpha}$ with mean $0<\bar{\alpha}^{\M / \Dip}<\infty$ and variance $\sigma^2 _{\alpha ^{\M /\Dip}}$.

A monopole-like active noise at  site $z$ along the chain, contributes with an entry on the diagonal  of the diffusion matrix $d_{z,z}$. By contrast, a dipole-like noise acts with completely anti-correlated forces on  the beads at positions $z$ and $z+1$,  contributing with four entries in the diffusion matrix: $\left[d_{z,z}=d_{z+1,z+1}= (k_{\rm B}/\gamma) \alpha ^D_z \right]$ and $\left[d_{z+1,z}=d_{z,z+1}=-(k_{\rm B}/\gamma) \alpha _z ^D \right] $. Therefore, the non-equilibrium part of the diffusion matrix will be a sum over monopole and dipole contributions of the form
    \centerline{
       \includegraphics[width=80mm]{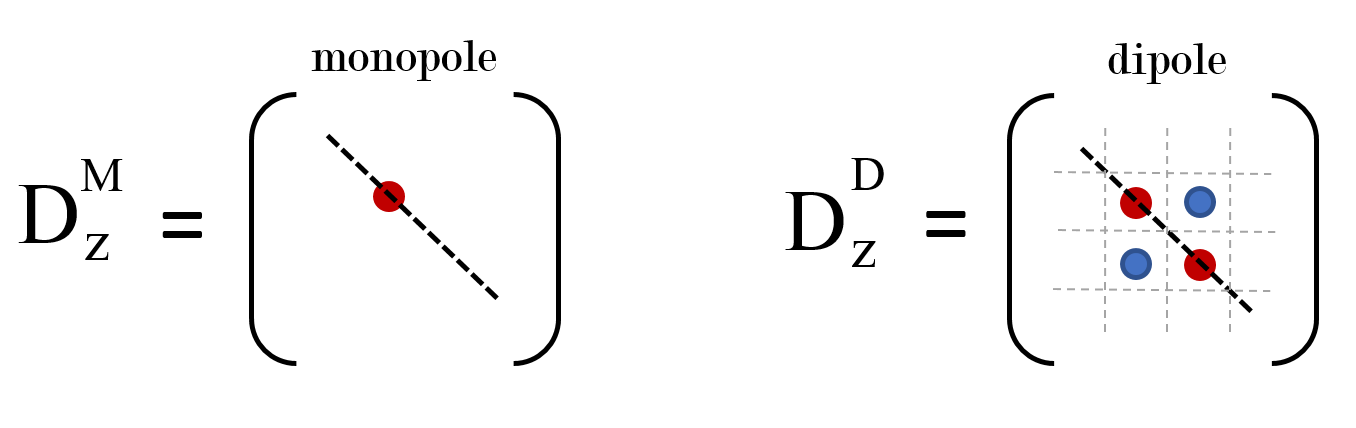}
    }

In general, owing to the linearity of the Lyapunov equation (see Eq.~\eqref{eq:Lyapunov}), the covariance matrix can be expressed as
\begin{equation}
\C=\frac{k_{\rm B}}{k}(T\eC + \sum _z \alpha_z ^{\M} \C _z^{\M} +\alpha_z ^{\Dip}\C _z^{\Dip}), 
\label{covariancesum}
\end{equation}
such that $\eC$, $\C _z^{\M}$ and $\C _z^{\Dip}$ are dimensionless. Here $\eC$ is the equilibrium solution of Eq.~\eqref{eq:Lyapunov} obtained for $\{ \alpha ^{{\M}/{\Dip}} _i\}=0$, and $\C _z^{{\M}/{\Dip}}$ are solutions of Eq.~\eqref{eq:Lyapunov} for single monopole/dipole noise of intensity $\alpha^{{\M}/{\Dip}} _z$ at position $z$  along the chain.
For a one-dimensional system the interaction matrix $\A$ has a simple form with entries: $a_{ij}=\frac{k}{\gamma}(-2 \delta _{ij} + \delta _{i,i+1} + \delta _{i,i-1})$. By inserting this expression for $a_{ij}$ together with Eq.~\eqref{covariancesum} into Eq.~\eqref{eq:AER}, we can express the area enclosing rate between any two tracer particles $i$ and $j$ that are not connected by a dipole activity ($d_{ij}=0$), as~\cite{Gradziuk2019}
\begin{equation}
\mathscr{A} _{ij}=\frac{k_{\rm B}}{\gamma}\sum _z (\xi_z ^{\M} \pdx c^{\M}_{z} + \xi_z ^{\Dip}\pdx c^{\Dip} _{z} ),
\label{eq:AER1d}
\end{equation}
where $\xi ^{{\M}/{\Dip}}=b^{{\M}/{\Dip}}\alpha ^{{\M}/{\Dip}}$ and we replaced the double indices in the coefficients $\{ b ^{\Dip}_{12},b ^{\Dip}_{23},\ldots,b ^{\Dip}_{i,i+1},\ldots,b ^{\Dip}_{N-1,N}\}$ with single index $\{ b ^{\Dip}_{1}, b ^{\Dip}_{2},\ldots, b ^{\Dip}_{i}, \ldots, b ^{\Dip}_{N-1}\}$. For notational simplicity we omit the subscripts $i,j$ on the right hand side, and we use $\pdx c =c_{i,j+1} -2c_{i,j} + c_{i,j-1}$ to indicate the discrete second derivative across rows. As can be seen from Eq.~\eqref{eq:AER1d}, the scaling behaviors of $\mathscr{A} _{ij}$, and therefore $\omega _{ij}$, are determined by the curvature of the covariance matrix. To find an expression for the curvature, we can calculate the explicit form of $C^{{\M}/{\Dip}}_{z}$, as a solution of the equation
\begin{equation}
\frac{k_{\rm B}\alpha ^{{\M}/{\Dip}}_z}{\gamma}\left(\pdy c^{{\M}/{\Dip}}_{z} +\pdx c^{{\M}/{\Dip}}_{z}\right)=-2d^{{\M}/{\Dip}}_{z},
\label{eq:discretediff}
\end{equation}
which is obtained from Eq.~\eqref{eq:Lyapunov} by considering a single monopole/dipole contribution at position $z$ along the chain. Eq.~\eqref{eq:discretediff} can be viewed as a discretized Poisson equation with sources set by the elements of $\D$.
The solution for a generic diffusion matrix will be obtained as a superposition of single-contribution solutions.
In the continuous limit, for a single entry in the diffusion matrix $d_{z,\tilde{z}}$, the single-source solution of Eq.~\eqref{eq:discretediff} gives $c(i,j)=-\frac{1}{\pi}\ln (\Varrho)+b$, where $\Varrho=\sqrt{(i-z)^2 + (j-\tilde{z})^2}$ indicates the distance from the source in the matrix and $b$ is an integration constant. Here we assumed radial symmetry of the solution around the source. 
The profile of $C^{\M/ \Dip}_{z}$ is different for the monopole and dipole cases and exhibits respectively a logarithmic profile $c^{\M}\propto \ln (\Varrho) $ and a power-law scaling $c^{\Dip} \propto \Varrho ^{-3}$, as illustrated in Fig.~\ref{fig:covsing}. These distinct scaling laws will later turn out to underlie the different scaling laws for the non-equilibrium measures. 

To determine the AER for a pair of beads at distance $r$ we need to compute contributions of the form
$\pdx c_{z;-r/2,r/2}^{\M/\Dip}\coloneqq \pdx c^{\M/\Dip}_{z}(r)$. Here for concreteness we have chosen beads with indices $i=-r/2$, $j=r/2$, where the index $0$ corresponds to the central bead. Note that the distance $r$ is dimensionless and measured in units of lattice spacing $\ell$. In the following we approximate the discrete derivatives applied to the matrix elements $\pdx c^{\M/\Dip}_{z}(r)$ with regular derivatives applied to the solutions of the Poisson equation $\pdc_2 c^{\M/\Dip}_{z}(r)$.

A single monopole activity at site $z$ contributes a diagonal entry in the diffusion matrix $d_{z,z}$. This appears as a monopole source at position $(z,z)$ in the Poisson equation. Using the continuous solution derived above we find~\cite{Gradziuk2019}
\begin{equation}
\pdc _{2}c^{\M}_{z}(r)=\frac{8rz}{\pi(r^2+4z^2)^2},
\label{eq:covmonopole}
\end{equation}

A single dipole activity between sites $z$ and $z+1$ contributes four entries in the diffusion matrix: $\left[d_{z,z}=-d_{z+1,z}=-d_{z,z+1}=d_{z+1,z+1}\right] $. These enter the Poisson equation as a quadrupole source. By summing these four contributions we obtain
\begin{align}
\begin{split}
\pdc _{2} c^{\Dip}_{z}(r)=&\frac{8rz}{\pi(r^2+4z^2)^2}\\
 &- \frac{4(-1+r)(1+2z)}{\pi(2+ (-2+r)r + 4z(1+z))^2}\\
&- \frac{4(1+r)(1+2z)}{\pi(2+(2+r)r + 4z(1+z))^2}\\
&+ \frac{8r(1+z)}{\pi(r^2+4(1+z)^2)^2}.
\label{eq:curvaturedipol}
\end{split}
\end{align}
From  Eq.~\eqref{eq:covmonopole} and Eq.~\eqref{eq:curvaturedipol}, it follows that when the only activity in the chain appears at one of the observed beads, for instance $z=i=-r/2$, then $\mathscr{A} _{ij}\sim r^{-2}$ for a monopole activity  and $\mathscr{A} _{ij} \sim r^{-5}$ for a dipole activity, in the limit $r\gg 1$.  As we will see below, the scaling of the curvature of the covariance (Eq.~\eqref{eq:covmonopole} and Eq.~\eqref{eq:curvaturedipol} underlies also the scaling behavior of the average quantities $\langle \omega^2(r) \rangle $ and $\langle {\mathscr{A}} ^2 (r) \rangle$ with distance.  

After this preparatory work, we are ready to obtain the first central results of this paper: Inserting Eq.~\eqref{eq:covmonopole} and Eq.~\eqref{eq:curvaturedipol} into Eq.~\eqref{eq:AER1d}, and averaging over configurations of the activity intensities, we obtain in the limit $r\gg 1$ (Sec.~(1) supplementary):
\begin{equation}
\label{eq:aerscaling}
\langle {\mathscr{A}} ^2 (r) \rangle _{\alpha}=\frac{1}{\pi}\left(\frac{k_{\rm B} }{\gamma} \right)^2 \left(\rho ^{\M}\sigma ^2 _{\alpha ^{\M}}\frac{1}{2r^3} + \rho ^{\Dip}\sigma ^2 _{\alpha ^{\Dip}}\frac{45}{r^7}\right).
\end{equation}

\begin{figure}
\centering
  \includegraphics[width=8.5cm]{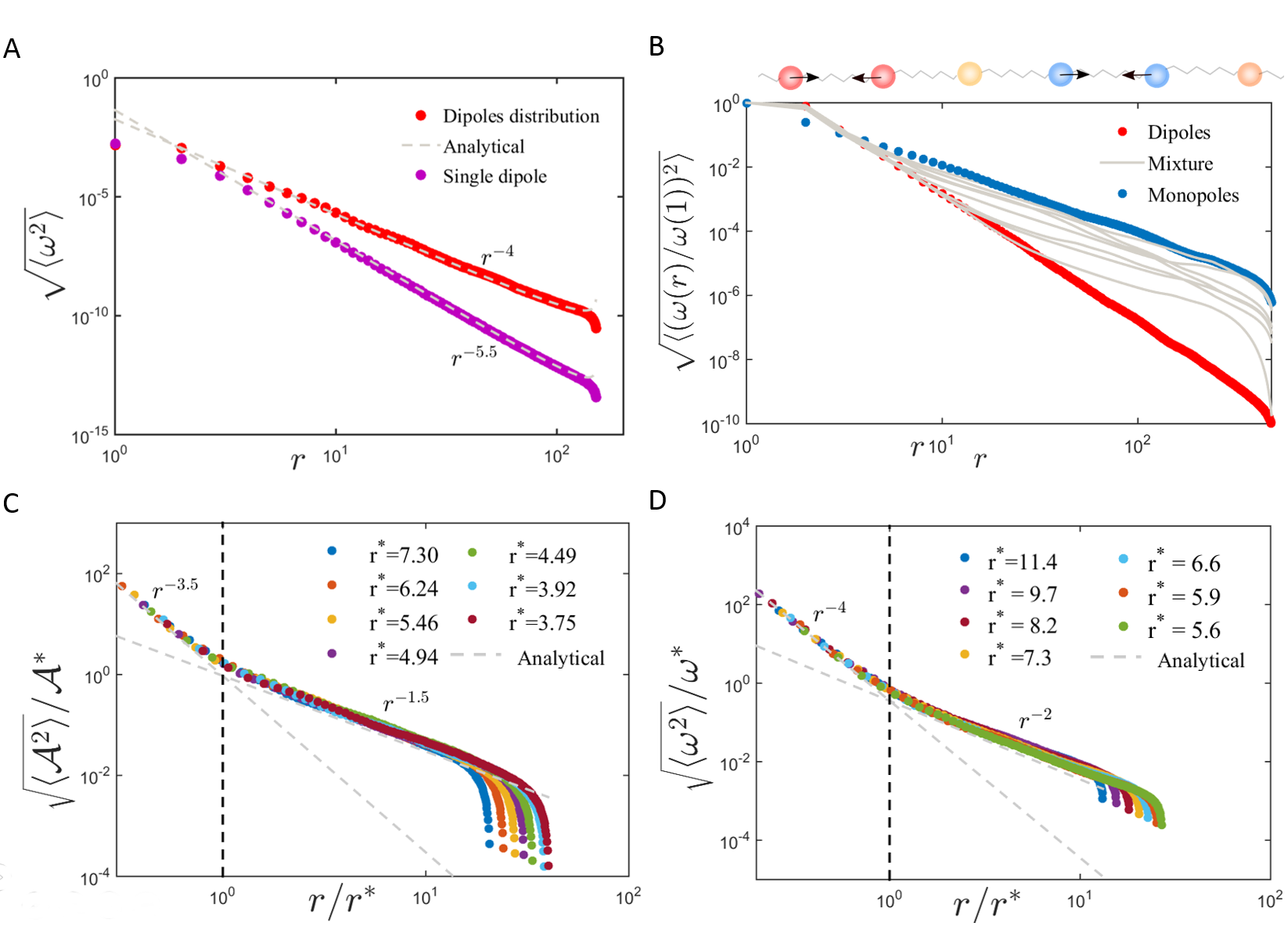}

  \caption{A) Scaling behavior of the cycling frequency with $r$ for a system with only dipoles as active noise. Numeric result in red and analytic prediction in gray. B) Scaling behavior of the cycling frequency with $r$ for  a system with activity given by: only dipoles (red), only monopoles (blue), mixture of dipoles and monopoles (gray) C) Rescaling of $\mathscr{A} $  for a range of $r^*$ and  $\mathscr{A} ^*$, and analytic prediction of the two scaling regimes in gray.  D) Rescaling of $\omega $ for a range of $r^*$ and $\omega ^*$, and analytic prediction of the two scaling regimes in gray.
}
    \label{fig:Fig2_ddipolescaling}
\end{figure}

Thus, we observe two different regimes: a monopole-controlled regime $ \langle {\mathscr{A}}  ^2 (r)\rangle _{\alpha} \propto 1/r^3 $ for long distances and a dipole-controlled regime  $\langle  {\mathscr{A}} ^2 (r)\rangle _{\alpha} \propto 1/r^7$ for short distances, as shown in Fig.~\ref{fig:Fig2_ddipolescaling}C.  Recall that $r$ is measured in units of the lattice spacing $\ell$. To rewrite Eq.~\ref{eq:aerscaling} in terms of the actual distance, $r \ell$, between the particles would require introducing a dependence on $\ell$ in both terms. Furthermore, we notice that in our model the dipole size equals $\ell$, and a generalization of the model to describe dipoles of arbitrary size would introduce a dependence on the dipole's size in the second term of Eq.~\ref{eq:aerscaling}

A similar scaling behavior can be obtained also for the cycling frequencies $\omega$ by applying a similar approach to Eq.~\eqref{eq:frequency}: Decomposing $\C= \frac{k_B}{k}\left(T\eC + \bar{\alpha} ^{\M}\C^{\M} + \bar{\alpha} ^{\Dip}\C ^{\Dip}\right) $  and expanding the factor $1/(\text{det}\C)$ up to linear order in $\frac{\bar{\alpha}^{\M / \Dip}}{T}$, we obtain
\begin{equation}\label{eq:omegascaling}
\langle \omega ^2 (r)\rangle _{\alpha}= \frac{(k/\gamma)^2}{\pi T^2\text{det}\eCr}\left(\rho ^{\M}\sigma ^2 _{\alpha^{\M}}\frac{1}{2r^3} + \rho ^{\Dip}\sigma ^2 _{\alpha ^{\Dip}}\frac{45 }{r^7}   \right).
\end{equation}
Considering  that~\cite{Gradziuk2019} for $r\ll N$, $\text{det}\eCr \propto Nr $,  we obtain two different regimes also for the scaling of the cycling frequency: $ \langle \omega  ^2(r)\rangle _{\alpha} \propto 1/r^4 $ at short distances and $ \langle \omega  ^2(r)\rangle _{\alpha} \propto 1/r^8 $ at long distances (Fig.~\ref{fig:Fig2_ddipolescaling}B,D).
Interestingly, in the presence of a distribution of only dipoles in the systems ($\rho ^{\M} =0$), the scaling exponent for the average cycling frequency  ($\sqrt{\langle \omega ^2 \rangle _{\alpha}} \sim r^{-4}$) differs from the exponent obtained with a single dipole activity at site $z=i=-r/2$ ($\sqrt{ \omega ^2 }\sim r^{-5.5}$) (Fig.~\ref{fig:Fig2_ddipolescaling}A). The same holds for a distribution of only monopoles ($\rho ^{\Dip} =0$), for which the scaling exponent $\sqrt{\langle \omega ^2 \rangle _{\alpha}}\sim r^{-2}$ differs from the single monopole scaling ($\sqrt{ \omega ^2 }\sim r^{-2.5}$)~\cite{Gradziuk2019}.  

Which parameters of our model determine the crossover distance $r^*$ between the two scaling regimes?
A direct calculation of $r^*$ and the mean cycling frequency at the crossover distance $\langle \omega ^2(r^*) \rangle$  leads to
\begin{align}
r^*=\left(\frac{90 \rho ^{\Dip}\sigma ^2 _{\alpha ^{\Dip}}}{\rho ^{\M}\sigma ^2 _{\alpha ^{\M}}} \right)^{1/4} && \langle\omega ^2(r^*) \rangle _{\alpha}= \frac{2k^2}{\pi T^2\gamma^245N}\frac{{\rho^{\M}} ^2 \sigma ^4 _{\alpha ^{\M}}}{\rho ^{\Dip} \sigma ^2 _{\alpha ^{\Dip}}}.
\end{align}
Note that $\langle \omega^2\rangle$ decreases as $1/N$, due to dependence of the determinant (see Eq.~\eqref{eq:omegascaling}) on system size: $\det \eCr \sim  N$. However, such a scaling with  system size is a property of one-dimensional systems, and will not appear in $d=2$ and $d=3$, where we expect respectively $\det \eCr \sim \ln (N)$ and $\det \eCr \sim \text{constant}$~\cite{Gradziuk2019}.

For the the area enclosing rate at the crossover point, we find
\begin{equation}
\langle\mathscr{A} ^2(r^*) \rangle _{\alpha}=\frac{k_{\rm B}^2}{\pi\gamma^290}\frac{{\rho^{\M}} ^2 \sigma ^4 _{\alpha ^{\M}}}{\rho ^{\Dip} \sigma ^2 _{\alpha ^{\Dip}}}r^*,
\end{equation}
as shown in Fig.~\ref{fig:Fig2_ddipolescaling}C,D, where we obtain a collapse of data by rescaling the x-axes by $r^*$ and the y-axes by $\mathscr{A^*}$ and $\omega^*$. 

Our results provide an indication of what kind of measurements could be performed to gain information on the non-equilibrium driving in an elastic system. For instance, the experimental observation of one of the scaling behaviors discussed here, e.g. for the cycling frequencies or area enclosing rates, would allow one to discriminate the monopole or dipole nature of the active driving. In the more general case of a mixture of monopoles and dipoles, a direct measure of the transition points between two different scaling regimes would help to gain quantitative information on the quantities $ \rho ^{\Dip} \sigma ^2 _{\alpha ^{\Dip}}$ and $\rho ^{\M} \sigma ^2 _{\alpha ^{\M}}$.

\subsection{Two-dimensional network}
\label{sec2d}
\begin{figure}
\centering
  \includegraphics[width=8.5cm]{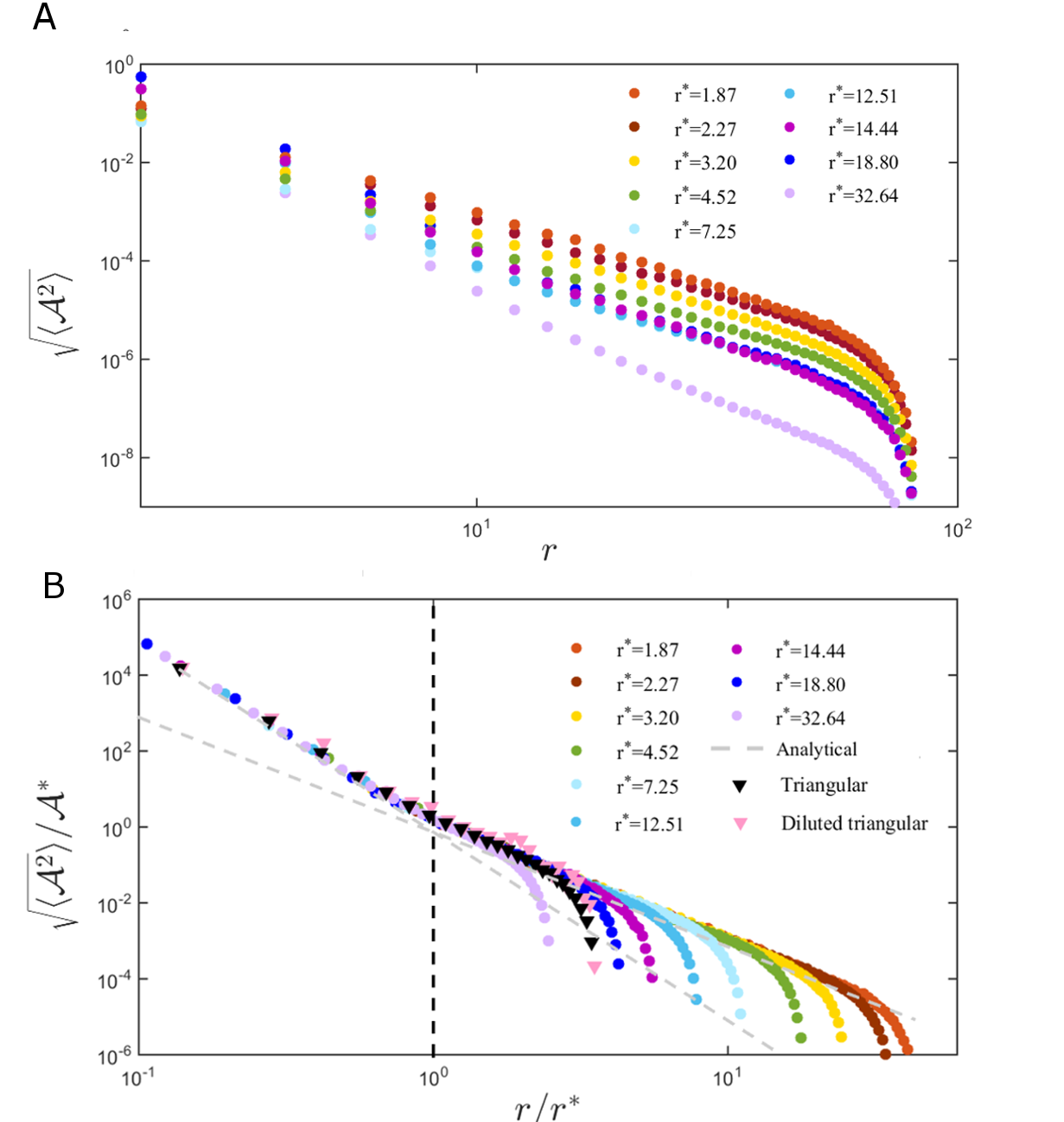}

  \caption{A)Scaling behavior of $\langle \mathscr{A} ^2 \rangle$ as function of the distance for two-dimensional lattices obtained for a square lattice and  different values of $r^*$. B) Data-collapse of the results in  A, obtained by rescaling the $x$ and $y$ axes respectively by $r^*_{2d}$ and ${\mathscr{A}({r_{2d}}^*)} ^2 $. The triangular markers correspond respectively to the data for a full triangular lattice (in black) and a randomly diluted triangular lattice (in pink) with dilution probability $p_d=0.8$
}
    \label{fig:scaling_2d}
\end{figure}
In this section we discuss how the results from Sec.~(\ref{sec1d}) can be extended to the case $d>1$. For simplicity, we focus on a two-dimensional square network, but we will discuss how the results obtained for this simple case also apply to more complex geometries. In addition, we consider the dipole forces acting always along the principal axes of the network, which corresponds to the limit of small displacements in the systems (Sec.~(2) supplementary).

We denote the elements of the covariance matrix, corresponding to beads at sites $\iv=(i_x,i_y)$ and $\jv=(j_x,j_y)$ in the lattice, as $c_{i_x,i_y;j_x,j_y}$. 
We consider zero rest length springs in such a way that the $x$ and $y$ coordinates decouple. Therefore, by $\C$ we mean the covariance matrix of only the degrees of freedom that correspond to a single chosen direction, and we can restrict the dipoles forces to always act along such direction, for instance the $y$-direction. This allows us to have employ a  one to one correspondence between dipoles and sites of the network.
From Eq.~\eqref{eq:AER}, we find~\cite{Gradziuk2019}
\begin{equation}
\mathscr{A}_{i_x,i_y;j_x,j_y}=\frac{k}{\gamma} \sum _{l=1} ^2 {\tilde{\partial} ^2} _l  c_{i_x,i_y;j_x,j_y},
\end{equation}
for beads not connected by a dipole activity ($d_{i_x,i_y;j_x,j_y}=0$). Here, the index $l$ runs over the directions $i_x$ and $i_y$. As in the one-dimensional case, we have a direct relation between  $\mathscr{A}$ and the second derivatives of the covariance matrix. Therefore, to find the scaling of $\mathscr{A}$ with the distance between two beads $r$, we need to find an expression for ${\tilde{\partial} ^2} _l  c_{i_x,i_y;j_x,j_y} (r)$.

As in Sec.~(\ref{sec1d}), we can rewrite the covariance matrix as: $\C=\frac{k_{\rm B}}{k}\left( T\eC + \sum _{\z}{\xi _{\z}^{\M} \C _{\z}^{\M} +\xi _{\z} ^{\Dip} \C _{\z} ^{\Dip} } \right)$, where we summed over contributions from all the monopole and dipole activities at sites $\z=(z_x,z_y)$ in the lattice. 
The continuous limit of the Lyapunov equation (Eq.~\eqref{eq:Lyapunov}) gives a Poisson equation in a four-dimensional space~\cite{Gradziuk2019}. In the case of a single non-zero entry in the diffusion matrix, $d_{{z_x,z_y};{\tilde{z}_x,\tilde{z}_y}}$, solving the Poisson equation gives $c(\iv,\jv)\sim \Varrho^{-2}$, where $\Varrho=\sqrt{(\iv-\z)^2+(\jv-\tilde{\z})^2}$ is the distance from the source. Using this continuous solution and replacing the discretized derivatives with standard ones we find
\begin{equation}
\partial^2 c(\iv,\jv)= \frac{2}{\pi^2} \frac{(\iv-\z)^2  - (\jv-\tilde{\z})^2}{((\iv-\z)^2  + (\jv-\tilde{\z})^2)^3},
\end{equation}
where we redefined $\partial^2=\sum _{l=1} ^2 {\partial} ^2 _l$.

A monopole activity at site $\z=(z_x,z_y)$ contributes a diagonal entry in the diffusion matrix: $d_{{z_x,z_y;{z_x},{z_y}}}$. As two beads at distance $r$ we can take, for instance, $(i_x,i_y)=(0,r/2)$ and $(j_x,j_y)=(0,-r/2)$, where for convenience we index the beads with the network center at $\left(0,0\right)$. Then, for the case of a single monopole activity at the site $(z_x,z_y)$ we obtain
\begin{equation}
\partial^2 c_{\z}^{\M}(r)= \frac{2}{\pi^2} \frac{16rz_y}{ (r^2+4(z_x^2+z_y^2))^3}.
\end{equation}
If we now consider a dipole random force of intensity $\alpha _i$ between two neighboring beads at position $(z_x,z_y)$ and $(z_x,z_y+1)$, this would correspond to four non-zero entries in the diffusion matrix: [$d_{z_x,z_y;z_x,z_y}=d_{z_x,z_y+1;z_x,z_y+1}=-d_{z_x,z_y;z_x,z_y+1}=-d_{z_x,z_y+1;z_x,z_y}$].
Therefore a single dipole activity enters the Poisson equation as a quadrupole source. Summing over all four contributions, we obtain for the second derivative of the covariance matrix for a single dipole activity
\begin{align}
\begin{split}
\partial^2 c_{\z}^{\Dip}(r)=& \frac{2}{\pi^2} \frac{8(1-r)(1+2z_y)}{ (2+(-2+r)r +4z_x^2+4z_y(1+z_y))^3} \\
&-\frac{2}{\pi^2}\frac{8(1+r)(1+2z_y)}{  (2+(2+r)r +4z_x^2+4z_y(1+z_y))^3} \\
& + \frac{2}{\pi^2} \frac{16rz_y}{ (r^2+4(z_x^2+z_y^2))^3}\\
&  +\frac{2}{\pi^2}\frac{16r(1+z_y)}{ (r^2+4(z_x^2+(z_y+1)^2))^3},
\end{split}
\end{align}
With similar steps as in the one-dimensional case, we find for the area enclosing rate: 
\begin{equation}
\label{AER2d2}
\langle \mathscr{A} ^2 (r)\rangle _{\alpha} \simeq \left(\frac{2k_{\rm B}}{\pi^2\gamma}\right) ^2\left[\rho^{\M} \sigma ^2 _{{\alpha}^{\M}} \frac{2\pi}{5r^6}+ \rho^{\Dip} \sigma ^2 _{{\alpha}^{\Dip}} \frac{529}{r^{10}}  \right].
\end{equation}
where to obtain the numerical prefactor and the scaling exponent in the second term we used a linear interpolation (Sec.~(1) supplementary).
Similarly to the one-dimensional case (see Eq.~\eqref{eq:aerscaling}), we recognize the presence of two different regimes at short and long distances. The crossover distance ${r^*}_{2D}$  between the monopole and dipole dominated regimes and the corresponding $\langle {\mathscr{A}^2({r^*_{2D}})}  \rangle$ read:
\begin{align}
\label{crossover2d}
\begin{split}
r^*_{2D}&\simeq 4.5\left(\frac{\rho^{\Dip} \sigma ^2 _{{\alpha}^{\Dip}}}{\rho^{\M} \sigma ^2 _{{\alpha}^{\M}}}\right)^{1/4} ,\\ 
 \langle {\mathscr{A}^2({r^*_{2D}})}  \rangle _{\alpha} &\simeq \frac{k_{\rm B}^2 12 \cdot 10^{-6}}{\gamma ^2}\frac{\left(\rho^{\M} \sigma ^2 _{{\alpha}^{\M}}\right)^{5/2}}{(\rho^{\Dip}\sigma ^2 _{{\alpha}^{\Dip}})^{3/2}}.
\end{split}
\end{align}
These analytical results are confirmed by the numerical data in Fig.~\ref{fig:scaling_2d}A, B, where we obtain a collapse by properly rescaling the $x$-axis by $r^*$ and the $y$-axis by $\mathscr{A}^2({r_{2D}}^*)$.

To investigate the sensitivity of these results to the specific underlying lattice structure, we studied  triangular and bond-diluted triangular networks. We find numerically that such lattice geometries are also well described by Eqs.~\eqref{AER2d2} and \eqref{crossover2d}. In fact, the curves numerically obtained for the scaling behavior of $\langle  \mathscr{A} ^2 \rangle _{\alpha}$ in a triangular network and diluted triangular network, overlap with the curves corresponding to the square lattice, as shown in Fig.~\ref{fig:scaling_2d} B.

\subsection{Spatially correlated activities}
\label{sec_corr}
\begin{figure}
\centering
  \includegraphics[width=8.5cm]{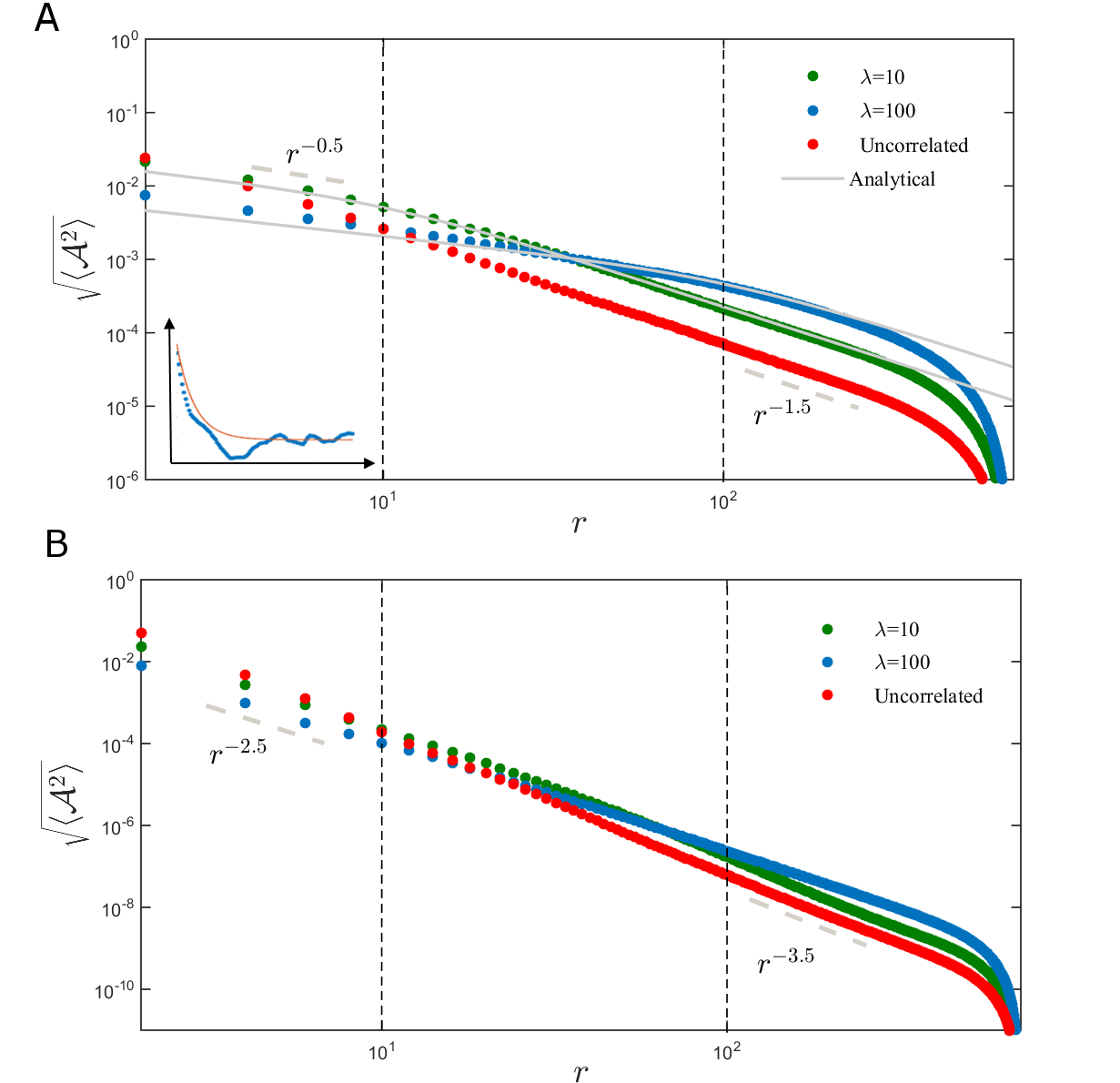}

  \caption{A) Scaling behavior of the area enclosing rate for a system with correlated monopole-activities with correlation length $\lambda =10$ (green), $\lambda =100$ (blue) and uncorrelated activities (red). In gray the comparison with the analytical prediction. In the inset: sketch of  the correlation function calculated for one realization of activities with correlation length $\lambda$ . B) Scaling behavior of the area enclosing rate for a system with dipole-activities  with $\lambda =10$ (green), $\lambda =100$ (blue) and uncorrelated activities (red). 
}
    \label{fig:Fig2_correlat}
\end{figure}
Up to this point, we considered two kinds of noise sources: dipoles and monopoles, randomly distributed in space. However, in biological systems the intensities of active processes may exhibit spatial correlation, for example due to the spacial organization of enzymes and molecular motors~\cite{Sutherland2009,Hachet2011,Schmitt2017}.  Therefore, it is crucial to determine how the spatial distribution of activities influences the scaling behavior of non-equilibrium measures. In this section, we consider a system where the intensities of the active noise are spatially correlated. 

As an illustrative example we consider a one-dimensional chain with only monopole activities ($\rho ^{\Dip} = 0$ and $\rho ^{\M} = 1$) and thus the diffusion matrix $\D$ is diagonal, with entries $d_{i,j}= (k_{\rm B}/\gamma)(T+\alpha _i) \delta _{i,j}$.  We draw the amplitudes $\{\alpha _i\}$ randomly from a probability distribution with covariance:  $\langle \alpha _i \alpha _j \rangle -  \langle \alpha _i\rangle \langle  \alpha _j \rangle =\sigma ^2 _{\alpha} e^{-\frac{|i-j|}{\lambda}} $, exhibiting a characteristic correlation length $\lambda > 0$ (see Fig.~\ref{fig:Fig2_correlat}A).
By setting $c^{\Dip}=0$ in Eq.~\eqref{eq:AER1d}, and considering the monopole contributions as in Eq.~\eqref{eq:covmonopole}, we obtain for the area enclosing rate:
\begin{align}
\begin{split}
&\langle {\mathscr{A} ^2 } (r)\rangle _{\alpha }=\left(\frac{k_{\rm B}}{\pi\gamma}\right)^2 \sum _{z,z'} \langle \alpha _z \alpha _{z'}\rangle  \frac{8rz}{(r^2+4z^2)^2} \frac{8rz'}{(r^2+4z'^2)^2}\\
&=\left(\frac{k_{\rm B}}{\pi\gamma}\right)^2 \sum _{z,z'} \sigma ^2 _{\alpha} e^{-\frac{|z-z'|}{\lambda}}  \frac{8rz}{(r^2+4z^2)^2}\frac{8rz'}{(r^2+4z'^2)^2}\\
&\simeq \left(\frac{ k_{\rm B}}{\pi\gamma}\right)^2 \sum _{z,z'=z-\lambda}^{z'=z+\lambda} (1- \frac{|z-z'|}{\lambda})  \frac{\sigma^2 _{\alpha}8rz}{(r^2+4z^2)^2} \frac{8rz'}{(r^2+4z'^2)^2},\\
\end{split}
\end{align}
where in the last step we approximated the exponential with a first order Taylor expansion inside the interval $[z-\lambda, z+\lambda ]$, and zero outside the interval. Approximating the sum with an integral, we arrive at:
\begin{equation}
\langle \mathscr{A}  ^2 (r) \rangle _{\alpha } \simeq \left(\frac{k_{\rm B}}{\gamma}\right)^2\frac{\sigma ^2 _{\alpha}}{\pi }\left(\frac{ \lambda}{2r^3 +2r\lambda^2}\right).
\label{eq:AER_correlated}
\end{equation}
We identify two different regimes: 
\[ \langle \mathscr{A}  ^2 (r) \rangle _{\alpha } \sim \begin{cases} \frac{k_{\rm B}^2 \sigma^2 _{\alpha} }{2\gamma ^2 \pi}\frac{\lambda}{r^3}& \mbox{if } r \gg \lambda \\  \frac{k_{\rm B}^2\sigma^2 _{\alpha}}{2\gamma ^2\pi }\frac{1}{\lambda r} & \mbox{if } r \ll\lambda \end{cases} \]
A comparison between our analytical prediction (Eq.~\eqref{eq:AER_correlated}) and numerical results is shown in Fig.~\ref{fig:Fig2_correlat}A, and the numerical results for correlated dipoles are shown in Fig.~\ref{fig:Fig2_correlat}B.
A similar calculation can be performed also for the cycling frequencies:

\[ \langle \omega (r)^2 \rangle _{\alpha } \sim \begin{cases} \frac{2 k^2\sigma^2 _{\alpha} }{T^2\gamma ^2\pi N}\frac{\lambda}{r^4} & \mbox{if } r \gg \lambda \\ \frac{2k^2\sigma^2 _{\alpha}}{T^2\gamma ^2\pi N}\frac{1}{\lambda r^2} & \mbox{if } r \ll\lambda \end{cases} \]
where we expanded the factor $\frac{1}{\text{det}\C}$ up to linear order in $\frac{\bar{\alpha}^{M}}{T}$ and replaced $\text{det}\eCr \simeq Nr/4$.

We can notice how the case of correlated intensities exhibits a behavior that is quantitatively different from the previous case of dipoles and monopole mixtures. In contrast to the previous case, the scaling exponent of the short distance regime is weaker than the one of the long distances regime. The exponent for the long distances ($r \gg \lambda $) is the same as for the case of uncorrelated activities. 

We summarized the scaling exponents obtained in $d=1$ for the area enclosing rate and cycling frequencies in Tab.~(\ref{table}).
\begin{table}

\begin{center}
\begin{tabular}
{c|ccc} &single  & $r<\lambda $ & $r>\lambda$\\
\hline 
\rule{0pt}{4ex}  
\includegraphics[width=0.5cm]{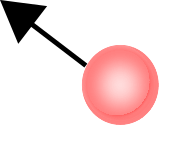} & 2.5 (2) & 1 (0.5) &2 (1.5)\\
\includegraphics[width=1cm]{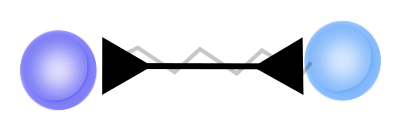} &5.5 (5) & 3 (2.5)& 4 (3.5)  \\
\end{tabular}
  \caption{Summary of the scaling exponents of cycling frequency $ \sqrt{\langle \omega (r)^2 \rangle _{\alpha } }$, and area enclosing rate $\sqrt{\langle \mathscr{A}  ^2 (r) \rangle _{\alpha } }$ (in brackets) obtained in $d=1$ for different cases: in the first column the result for the case of a single monopole and dipole source, in the second and third columns the results for the monopoles and dipoles distributions, obtained for distances smaller and bigger then the correlation length $\lambda$.
\label{table}}
\end{center}
\end{table}
The differences in the observed scaling exponents would allow one to discern the cases of spatially correlated activities and of mixture of dipoles and monopoles. Furthermore, a quantitative measure of the two scaling regimes, would allow one to estimate the correlation length $\lambda$ or the intensities of the activities $\sigma^2 _{\alpha} $. 

\section{Conclusions}
In this work we asked how the scaling behavior of two-point non-equilibrium measures can be used to reveal properties of the internal driving in an active elastic assembly. To this end, 
 we considered a lattice model of a driven elastic assembly. Using this model, we investigated how intrinsic features of the active noise influence the scaling behavior of non-equilibrium measures, such as cycling frequencies $\omega$ and area enclosing rates ${\mathscr{A}}$. These measures are directly accessible from the stochastic trajectories of pairs of tracer particles in the network. 

Using our theoretical framework, we considered several settings of the active noise. We started by focusing in Sec.~(\ref{sec1d}) on a one-dimensional system driven out of equilibrium by a mixture of stochastic monopole and dipole forces. We performed an analytical calculation to find an expression for the scaling law of $\sqrt{\langle \mathscr{A}^2 \rangle _{\alpha }}$ and $\sqrt{\langle \omega ^2 \rangle _{\alpha } }$ as a function of the distance between the observed particles, which we confirmed by numerics. We predict two scaling regimes: a dipole-dominated regime at short distances and a monopole-dominated regime at long distances. The crossover length between these two regimes is set by the parameters characterizing the stochastic forces: the densities of dipoles $\rho ^{\Dip}$ and monopoles $\rho ^{\M}$, and the variance of their intensities $\sigma^2_{\alpha^{\M/\Dip}}$.
We extended these results in Sec.~(\ref{sec2d}), where we performed analogous calculations for a two-dimensional network, and observed qualitatively the same behavior as for the one-dimensional system, but with different exponents. Importantly, we demonstrated numerically that our predictions, obtained for a square lattice, also apply to more complex networks such as triangular and diluted triangular networks, more commonly employed to describe soft biological materials~\cite{Broedersz2011a,Turlier2016,Gnesotto2018,Vahabi2016}. Since in real systems active noise amplitudes may be spatially correlated, in Sec.~(\ref{sec_corr}) we considered an illustrative example of a system driven out of equilibrium by stochastic forces with intensities correlated exponentially in space. Interestingly, we find that these correlations are reflected as a weaker decay in the scaling behavior of our non-equilibrium measures on lengthscales below the correlation length (see Tab.~(\ref{table})).

Altogether our results provide a new perspective to interpret experimentally accessible two-point non-equilibrium measures: A direct observation of the scaling behavior of such non-equilibrium measures may provide a way to infer qualitative information on the nature of the active forces in the system, and quantitative information on their densities, intensities, or their correlation length. A typical setting where our approach could be applied is  time-lapse microscopy experiments in which several probe particles are tracked in active assemblies of soft materials~\cite{balland2006power, Lau2003}. Promising examples would be \textit{in vitro} or  \textit{in vivo} biological actomyosin networks, cellular membranes, DNA polymers, but also synthetic and biomimetic systems~\cite{Gnesottoreview, Needleman2017b,Palacci2013,Bertrand2012,Manneville2001,Manneville1999}. Our approach could help connect mesoscale non-equilibrium dynamics to the microscopic properties of the internal driving in such systems.

\section*{Acknowledgements}
We thank F. Gnesotto, S. Ceolin, B. Remlein, and G. Torregrosa Cortes for many stimulating discussions.
This work was supported by the German Excellence Initiative via the program NanoSystems Initiative Munich (NIM), the Graduate School of Quantitative Biosciences Munich (QBM),  and was funded by the Deutsche Forschungsgemeinshaft (DFG, German Research Foundation) - 418389167.\\

\bibliography{Newbiblio15}

\onecolumngrid
\section*{Supplementary Notes}
\beginsupplement
\subsection*{Derivation of $\langle \mathscr{A}^2(r) \rangle _{\alpha}$ in $d=1$ and $d=2$}
In this section we derive an expression for the average area enclosing rate $\langle \mathscr{A} ^2 \rangle $ as function of the distance $r$ between two observed probes. For a one-dimensional system the area enclosing rate can be expressed in terms of the elements of the covariance matrix as:
\begin{equation}
\mathscr{A} _{ij}=\frac{k_B}{\gamma}\sum _z \left(\xi_z ^{\M} \pdx c^{\M}_{z} + \xi_z ^D\pdx c^{\Dip} _{z} \right),
\label{eq:AER1dSUP}
\end{equation}
where $i$ and $j$ are the bead indices such, that $d_{ij}=0$, $\pdx c =c_{i,j+1} -2c_{i,j} + c_{i,j-1}$ indicates the discrete second derivative across rows, and $\xi ^{{\M}/{\Dip}}=b^{{\M}/{\Dip}}\alpha ^{{\M}/{\Dip}}$.
To find how $\langle\mathscr{A}^2(r)\rangle _{\alpha}$ depends on the distance $r$ between the observed probes, we use the explicit expressions for $\pdc c^{\M}_{z}(r)$ and $\pdc c^{\Dip} _{z}(r)$, evaluated for $i=-r/2$ and $j=r/2$, appearing in Eq. (12) and Eq. (13) in the main text.

\begin{figure}[H]
\centering
  \includegraphics[width=170mm]{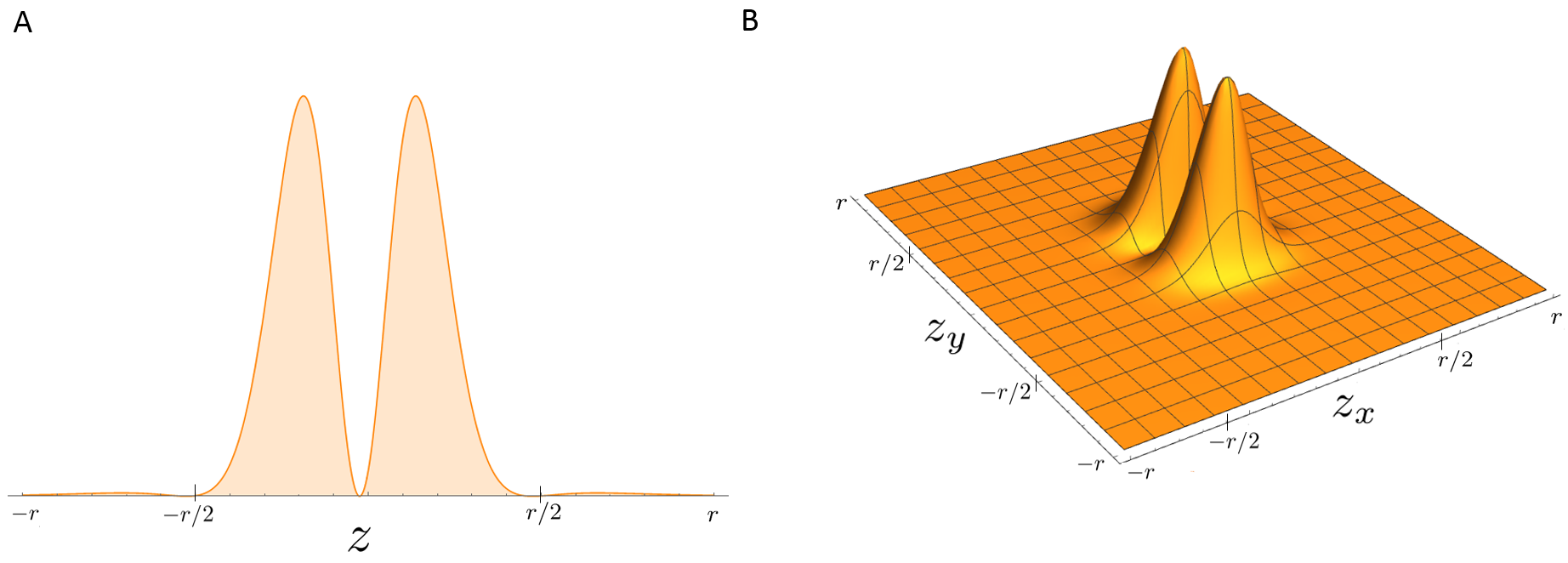}

  \caption{A) Plot of ${f^{\Dip}}^2(z)$ in $d=1$. The value of this function represents the contribution of a dipole, at position $z$, to the area enclosing rate measured between two tracers beads at $i=r/2$ and $j=-r/2$ B) Plot of ${f^{\Dip}}^2(z_x,z_y)$ in $d=2$. The value of this function represents the contribution of a dipole activity, acting between the beads at $\{z_x,z_y \}$ and $\{z_x,z_{y}+1\}$, to the area enclosing rate measured from the $y$-displacements of two beads at $\{0,r/2\}$ and $\{0,-r/2\}$. 
}
    \label{fig:Fig1_SUP}
\end{figure}

For notational simplicity we  rename: $\pdc c^{\M}_{z}(r)= f^{\M}(z,r)$ and $\pdc c^{\Dip}_{z}(r)= f^{\Dip}(z,r)$. The functions  $f^{\M}(z,r)$ and $f^{\Dip}(z,r)$ are informative of the contribution of a monopole or dipole activity, at position $z$, to $\mathscr{A}^2(r)$ measured between two tracers at $i=r/2$ and $j=-r/2$. Such contributions come primarily from activities in between the two beads, as shown in Fig.~\ref{fig:Fig1_SUP}. 

\begin{figure}
\centering
  \includegraphics[width=80mm]{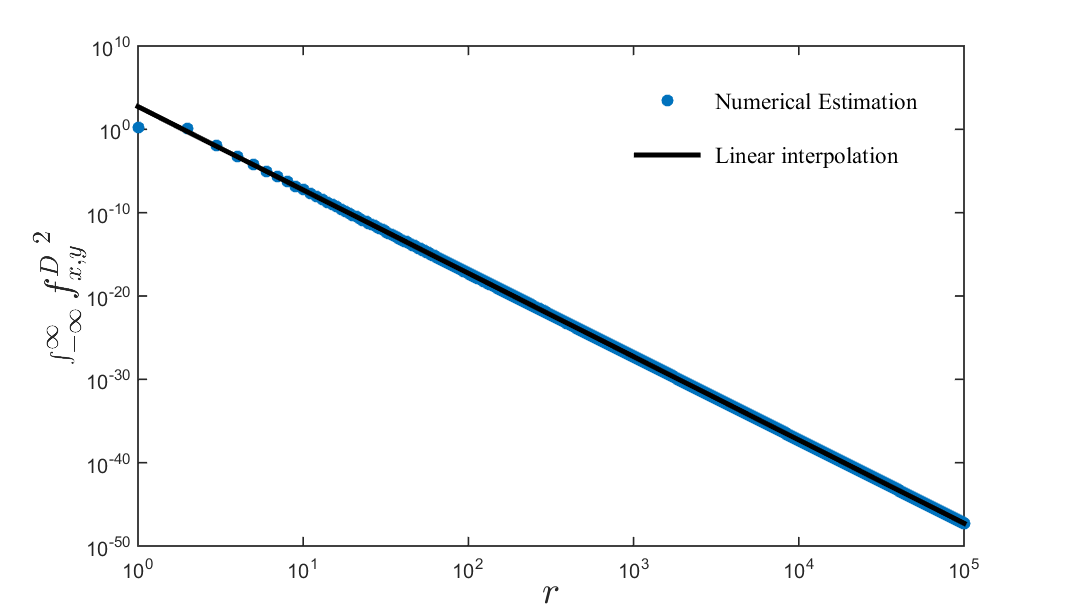}

  \caption{Numerical estimation of the second integral in Eq.~\eqref{AER2d} as function of the distance (blue), and  result of the linear interpolation $f(x)=6,27 -9.99 \ln (r)$ (black). 
}
    \label{fig:linear interpolation}
\end{figure}

By taking the square of Eq. \eqref{eq:AER1dSUP} and the ensemble average over the activities we obtain:
\begin{align}
\label{eq:AER1d2_1}
\begin{split}
\langle \mathscr{A} ^2(r) \rangle  _{\alpha} &=\frac{k_B^2}{\gamma ^2}\left\langle  \left[\sum _z \xi_z ^{\M} f^{\M}(z,r) + \xi_z ^{\Dip}f^{\Dip}(z,r) \right]^2 \right\rangle \\
&=\frac{k_B^2}{\gamma ^2} \sum _z \left[ \langle {\xi_z ^{\M}}^2\rangle  {f^{\M}}^2(z,r) + \langle {\xi_z ^{\Dip} }^2{\rangle f^{\Dip}}^2(z,r) \right] 
 + \frac{k_B^2}{\gamma ^2}\langle\xi^{\M}\rangle^2\sum _{z,z' \neq z}  f^{\M}(z,r)f^{\M}(z',r)\\
 &+\frac{k_B^2}{\gamma ^2} \langle\xi ^{\Dip}\rangle ^2 \sum_ {z,z' \neq z}  f^{\Dip}(z,r) f^{\Dip}(z',r)
 + \frac{k_B^2}{\gamma ^2}2\langle\xi ^{\M}  \rangle \langle\xi^{\Dip}\rangle\sum_ {z}  f^{\M}(z,r)\sum_ {z'} f^{\Dip}(z',r).\\
\end{split}
\end{align}
where we assumed that the noise amplitudes $\xi_z ^{\M}  $ and $\xi_{z'}^{\Dip}$  are spatially uncorrelated and that their average does not depend on $z$. By rewriting $\sum _{z,z'\ne z}=\sum _{z,z'} - \sum _{z,z'=z}$ we obtain:
 \begin{align}
\label{eq:AER1d2_2}
\begin{split}
\langle \mathscr{A} ^2(r) \rangle  _{\alpha} &=\frac{k_B^2}{\gamma ^2} \sum _z \left[ \langle {\xi_z ^{\M}}^2\rangle  {f^{\M}}^2(z,r) + \langle {\xi_z ^{\Dip} }^2{\rangle f^{\Dip}}^2(z,r) \right] 
 + \frac{k_B^2}{\gamma ^2}\langle\xi^{\M}\rangle^2\left[\sum _{z,z' }  f^{\M}(z,r)f^{\M}(z',r) - \sum _{z}  {f^{\M}}^2(z,r) \right] \\
 &+\frac{k_B^2}{\gamma ^2} \langle\xi ^{\Dip}\rangle ^2 \left[ \sum_ {z,z'}  f^{\Dip}(z,r) f^{\Dip}(z',r)- \sum_ {z}  {f^{\Dip}}^2(z,r)\right]
 + \frac{k_B^2}{\gamma ^2}2\langle\xi ^{\M}  \rangle \langle\xi^{\Dip}\rangle\sum_ {z}  f^{\M}(z,r)\sum_ {z'} f^{\Dip}(z',r).\\
\end{split}
\end{align}
and approximating the sum by an integral yields
 \begin{align}
\label{eq:AER1d2_3}
\begin{split}
\langle \mathscr{A} ^2(r) \rangle  _{\alpha} &=\frac{k_B^2}{\gamma ^2} \int dz \left[ \langle {\xi_z ^{\M}}^2\rangle  {f^{\M}}^2(z,r) + \langle {\xi_z ^{\Dip} }^2{\rangle f^{\Dip}}^2(z,r) \right] 
 + \frac{k_B^2}{\gamma ^2}\langle\xi^{\M}\rangle^2\left[\int dz dz' f^{\M}(z,r)f^{\M}(z',r) - \int dz {f^{\M}}^2(z,r) \right] \\
 &+\frac{k_B^2}{\gamma ^2} \langle\xi ^{\Dip}\rangle ^2 \left[ \int dz dz' f^{\Dip}(z,r) f^{\Dip}(z',r)- \int dz  f^{\Dip}(z,r)^2\right]
 + \frac{k_B^2}{\gamma ^2}2\langle\xi ^{\M}  \rangle \langle\xi^{\Dip}\rangle\int dz f^{\M}(z,r)\int dz' f^{\Dip}(z',r).\\
\end{split}
\end{align}
Considering that $\int f^{\M /\Dip}(z,r)dz =0 $:
\begin{equation}
\langle {\mathscr{A} ^2(r)} \rangle  _{\alpha}=\left(\frac{k_B}{\gamma} \right)^2 \left(\sigma ^2 _{\xi ^{\M}} \int dz  {f^{\M}}^2(z,r)+ \sigma ^2 _{\xi^{\Dip}}\int dz {f^{\Dip}}^2(z,r)\right). 
\end{equation}
Since $\sigma ^2 _{\xi ^{{\M}/{\Dip}}}$ is the variance of a stochastic variable $\xi ^{{\M}/{\Dip}}=b^{{\M}/{\Dip}}\alpha^{{\M}/{\Dip}}$, where $b \in \{0,1\}$ with $\langle b\rangle= \rho ^{{\M}/{\Dip}}  $ and $\langle b^2\rangle =\rho ^{{\M}/{\Dip}} $, we have $\sigma ^2 _{\xi ^{{\M}/{\Dip}}}=\rho ^{{\M}/{\Dip}}\sigma ^2 _{\alpha ^{{\M}/{\Dip}}}$.
By solving the integral, and keeping only the leading terms in the limit $r \gg 1$, we obtain
\begin{equation}
\langle {\mathscr{A}} ^2 (r) \rangle _{\alpha}=\frac{1}{\pi}\left(\frac{k_B}{\gamma} \right)^2 \left(\rho ^{\M}\sigma ^2 _{\alpha ^{\M}}\frac{1}{2r^3} + \rho ^D\sigma ^2 _{\alpha ^{D}}\frac{45}{r^7}\right).
\end{equation}
Similar calculations can be performed in d=2, and lead to the integral form of the area enclosing rate:
\begin{equation}
\label{AER2d}
\langle \mathscr{A}  ^2(r) \rangle _{\alpha} = \frac{k_B^2}{\gamma ^2}\left[\rho^{\M} {\sigma ^2 _{{\alpha}^{\M}}}\int _{-\infty} ^{\infty}dz_x dz_y {f^{\M} }^2(z_x,z_y,r)+ \rho^{\Dip}{\sigma ^2 _{{\alpha}^D}}\int _{-\infty} ^{\infty}dz_x dz_y {f^{D}}^2 (z_x,z_y,r) \right].
\end{equation}
where $f^{\M}(z_x,z_y,r)$ and  $f^{\Dip}(z_x,z_y,r)$ are defined in Eq. (20) and Eq. (21) in the main text. The second integral in Eq.~\eqref{AER2d} is arduous to calculate analytically. Therefore, we estimate the integral numerically for different values of the distance $r$. The result is reported in Fig.~\ref{fig:linear interpolation}, together with the result of a linear interpolation of such numerical data $\int _{-\infty} ^{\infty}dz_x dz_y {{f^{\Dip}}^2(z_x,z_y,r)} \simeq \frac{4}{\pi ^4} ar^{-b}$ with $a\simeq 529$ and $b\simeq 10$. Finally, for $d=2$ we obtain :
\begin{equation}
\label{AER2d2_SUP}
\langle \mathscr{A} ^2 (r)\rangle _{\alpha} \simeq \left(\frac{2k_{\rm B}}{\pi^2\gamma}\right) ^2\left[\rho^{\M} \sigma ^2 _{{\alpha}^{\M}} \frac{2\pi}{5r^6}+ \rho^{\Dip} \sigma ^2 _{{\alpha}^{\Dip}} \frac{529}{r^{10}}  \right].
\end{equation}

\subsection*{Small displacement approximation in $d=2$}
In a two-dimensional network, we assume that the dipole forces act along the directions of the springs at each point in time. Here we show that in the limit of small displacements we can consider the action of the dipole forces to be directed along the principal axes of the network at rest, as done in the main text. 
For simplicity, let's consider a square lattice of size $N=n\times n$, with zero-rest length springs. We index the coordinates as $\x = \{x_1,\ldots,x_{N},y_1,\ldots,y_{N}\}$, and we assume the presence of only two dipole activities: one of intensity $\eta _{i,i+n}$ acting vertically between the beads of index  $i$ and $i+n$, and the other one of intensity $\eta _{i,i+1}$ acting horizontally  between the beads of index  $i$ and $i+1$. 
The Langevin equations for the $y$-displacements of the $i_{th}$ bead reads: 
\begin{align}
\begin{split}
\frac{dy _i}{dt}&=\frac{k}{\gamma}a_{i,l} y_l + \eta ^T _i + \eta _{i,i+n} ^{\Dip} \cos \left(\frac{\Delta x_{i,i+n}}{\ell}\right) + \eta _{i,i+1} ^{\Dip} \sin \left(\frac{\Delta y_{i,i+1}}{\ell}\right) \\ 
\end{split}
\end{align}
 where we defined $\Delta x_{i,i+n} =x_{i}- x_{i+n}$,  $\Delta y_{i,i+1} =y_{i}- y_{i+1}$  and $\ell$ is the lattice spacing. In the limit $\frac{\Delta x_{i,i+n}}{\ell}\ll 1$ and $\frac{\Delta y_{i,i+1}}{\ell}\ll 1$, the last term is negligible and the second last is $\simeq  \eta _{i,i+n} ^{\Dip}$. Therefore in the limit of small displacements we can consider the action of the dipole forces to be directed along the principal axes of the network. 

\begin{figure}[H]
\centering
  \includegraphics[width=95mm]{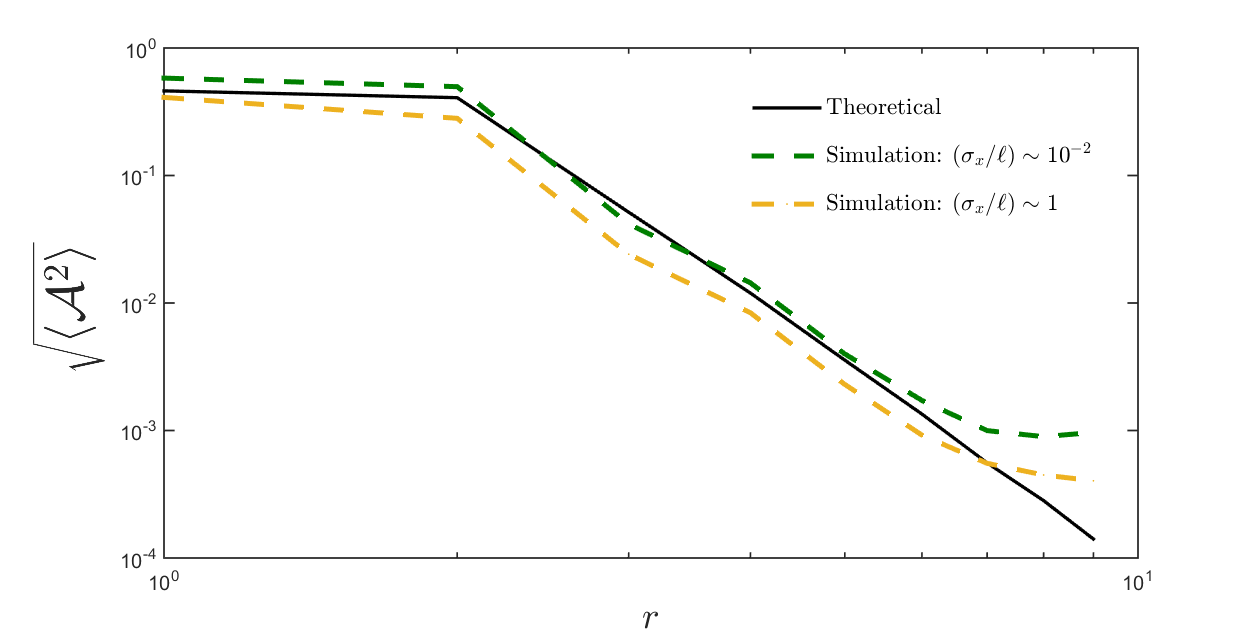}

  \caption{Theoretical prediction (black) of the average area enclosing rate (obtained from Eq.~7 in the main text by numerically solving the Lyapunov equation), compared with simulation results for different values of the ratio $\sigma _x/\ell $ (green and yellow). For computational convenience the ensemble average over the activity distributions has been evaluated by performing a spatial average over the lattice.
}
    \label{fig:comp_num_sim}
\end{figure}

To check the validity of this approximation for the non-equilibrium measure, we explicitly simulated the dynamics of the network. We employed the Euler-Maruyama method to numerically integrate the Langevin equation of a square lattice where both vertical and horizontal dipoles are distributed randomly along the network and act along the spring direction. When the standard deviation of displacements is small compared to the rest length of the springs ( $\sigma _{x}/\ell <1$), our theoretical prediction is in good agreement with the simulation, as shown in Fig.~\ref{fig:comp_num_sim}. However, also in the case of $\sigma _{x}/\ell >1$, the simulation results are slightly shifted respect to our prediction, but the scaling exponent remains the same.

\end{document}